\documentstyle[12pt,amssymb,epsfig]{article}

\everyjob{\typeout{MuTeX.sty: LaTeX Modification by EPM.}}
\immediate\write10{MuTeX.sty: LaTeX Modification by EPM.}

\makeatletter   

\newtheorem {Canham threshold} [theorem] {Canham Threshold}

\def\theoremstyle#1#2{\def\@@theoremheadstyle{#1}
                      \def\@@theorembodystyle{#2}}
\def\@@theoremheadstyle{\sc}
\def\@@theorembodystyle{\rm}
\def\@begintheorem#1#2{\@@theorembodystyle 
                       \trivlist 
                       \item[\hskip 
                             \labelsep{\@@theoremheadstyle #1\ #2}]}
\def\@opargbegintheorem#1#2#3{\@@theorembodystyle 
                              \trivlist 
                               \item[\hskip 
                                  \labelsep{\@@theoremheadstyle #1\ #2\ (#3)}]}

 \def\@@pc{\bf}

 \newcommand {\pcodestyle}[1] {\def\@@pc{#1}}  

 {
 \def\PROGRAM           {{\@@pc program\ }}
 
 \def\PROCEDURE         {{\@@pc procedure\ }}
 \def\FUNCTION          {{\@@pc function\ }}
 \def\LOCAL             {{\@@pc local\ }}
 \def\GLOBAL            {{\@@pc global\ }}
 \def\RETURNS           {{\@@pc returns\ }}
 \def\RETURN            {{\@@pc return\ }}
 \def\BEGIN             {{\@@pc begin\ }}
 \def\END               {{\@@pc end\ }}
 \def\IF                        {{\@@pc if\ }}
 \def\THEN              {{\@@pc then\ }}
 \def\ELSE              {{\@@pc else\ }}
 \def\REPEAT            {{\@@pc repeat\ }}
 \def\UNTIL             {{\@@pc until\ }}
 \def\WHILE             {{\@@pc while\ }}
 \def\DO                        {{\@@pc do\ }}
 \def\FOR               {{\@@pc for\ }}
 \def\TO                        {{\@@pc to\ }}
 \def\DOWN              {{\@@pc down\ }}
 \def\NEXT              {{\@@pc next\ }}
 \begin{center}
  \begin{minipage}{.9\textwidth}
   \small
    \begin{tabbing}
 }%
 {  \end{tabbing}%
   \end{minipage}%
  \end{center}%
 }
\def \epf{ \hfill $\Box$ }

                           {\end{quotation}}

                         {  \end{minipage}%
                           \end{center}%
                          \end{description}%
                         }

                                { \end{minipage}
                                 \end{center}%
                                }

\def\thebibliography#1{\section*{References}\list
 {[\arabic{enumi}]}{\settowidth\labelwidth{[#1]}
 \leftmargin\labelwidth
 \advance\leftmargin\labelsep
 \usecounter{enumi}}
 \def\newblock{\hskip .11em plus .33em minus -.07em}
 \sloppy
 \sfcode`\.=1000\relax}

\def\section{\@startsection {section}{1}{\z@}{-3.25ex plus-1ex minus
    -.2ex}{1.5ex plus.2ex}{\reset@font\large\bf}}
\def\subsection{\@startsection{subsection}{2}{\z@}{-3.25ex plus-1ex
    minus-.2ex}{1.5ex plus.2ex}{\reset@font\large\bf}}

\newsavebox{\ProofSym}
\savebox{\ProofSym}{%
  \begin{picture}(10,10)
    \put(0,0){\framebox(9,9){}}
    \put(0,3){\framebox(6,6){}}
  \end{picture}}
\newcommand{\eop}{\hfill\usebox{\ProofSym}}

\theoremstyle{\sc}{\rm}

\begin{document}



\begin{center}
      
 {\huge\bf
On joint triangulations of two sets of points in the plane \footnote{The extended abstart of this
paper appeared in the Proceedings of India-Taiwan Conference on Discrete Mathematics, Taipei,  
pp. 34-43, 2009.}}\\
      ~\\
     ~\\
      ~\\
      ~\\

\end{center}

\mbox{\begin{minipage} [b] {2.7in}
\centerline{Ajit Arvind Diwan}
\centerline{Dept. of Computer Science and Engineering}
\centerline{Indian Institute of Technology Bombay}
\centerline{Mumbai 400076, India} 
\centerline{Email address: aad@cse.iitb.ac.in}
\end{minipage}}\hspace{0.6in}
\mbox{\begin{minipage} [b] {2.7in}
\centerline{Subir Kumar Ghosh}
\centerline{School of Computer Science}
\centerline{Tata Institute of Fundamental Research}
\centerline{Mumbai 400005, India} 
\centerline{Email address: ghosh@tifr.res.in}
\end{minipage}}
\vskip .4in

\mbox{\begin{minipage} [b] {2.7in}
\centerline{Partha Pratim Goswami}
\centerline{Institute of Radiophysics and Electronics}
\centerline{University of Calcutta}
\centerline{Kolkata 700009, India} 
\centerline{Email address: ppg.rpe@caluniv.ac.in}
\end{minipage}}\hspace{0.6in}
\mbox{\begin{minipage} [b] {2.7in}
\centerline{Andrzej Lingas}
\centerline{Department of Computer Science}
\centerline{Lund University}
\centerline{ Lund S-22100, Sweden } 
\centerline{Email address: Andrzej.Lingas@cs.lth.se}
\end{minipage}}

\vskip .2in

\begin{abstract}

In this paper, we establish two necessary conditions for a joint triangulation
of two sets of $n$ points in the plane and conjecture that they are sufficient.
We show that these necessary conditions can be tested in $O(n^3)$ time.
For the problem of a joint triangulation of two simple polygons of $n$ 
vertices, we propose an $O(n^3)$ time algorithm for constructing a
joint triangulation using dynamic programming.

\end{abstract}

\section{Introduction}

Let $S$ be a set of points in the plane. A triangulation of $S$ is a maximal set of 
line segments with endpoints in $S$ such that no two segments intersect in their 
interior. A triangulation of $S$ partitions the convex hull of $S$ into regions not 
containing points in $S$ that are bounded by triangles. Triangulating a set of 
unlabeled points 
in the plane under various constraints is a well studied problem in computational 
geometry~\cite{bkos-cgaa-97,ddgc-thc-97, ps-cg-85}. 

\medskip

\begin{figure}
\begin{center}
\centerline{\hbox{\psfig{figure=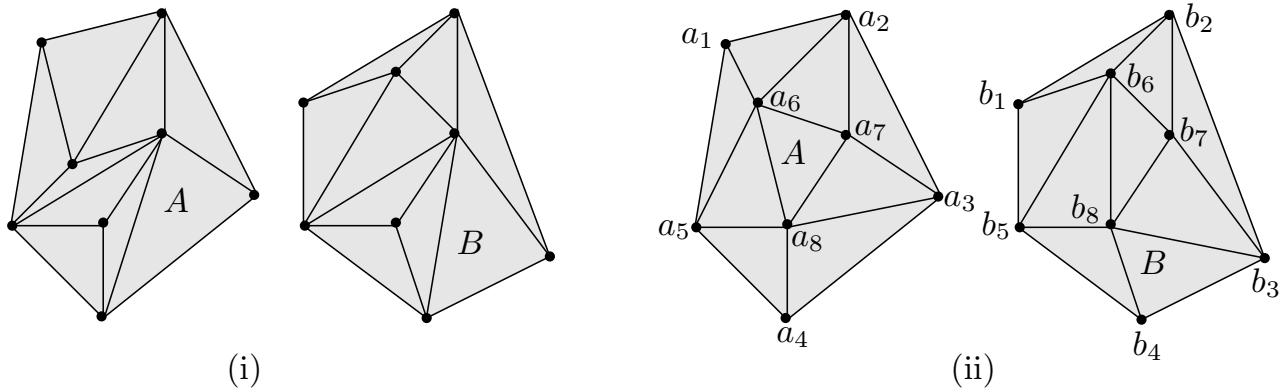,width=1.0\hsize}}}
\caption{Joint triangulations of two sets of points  $A$ and $B$: (i) bijection is not given, and (ii)
bijection is given.}
\label{ljoint-1}
\end{center}
\end{figure}

Consider two sets $A$ and $B$ of points in the plane, where $|A|=|B|=n$.
Two triangulations  $T_a$ of $A$ and $T_b$ of $B$ are called 
{\it joint triangulation} (also called {\it compatible triangulation})
of $A$ and $B$ if there exists a bijection $f$ between
$A$ and $B$  such that (i) $ijk$ is a triangle in $T_a$ 
if and only 
if $f(i)f(j)f(k)$ is a triangle in $T_b$, and (ii)  $ijk$ and $f(i)f(j)f(k)$
do not contain any point of $A$ and $B$ respectively (see Figure \ref{ljoint-1}).
The problem has applications in morphing \cite {sg-imct-2003, sg-hqct-2004} and automated cartography \cite{saal-jttm-87}.

\medskip
The problem of joint triangulation of   $A$ and $B$
has two variations depending upon whether  the bijection between points 
of $A$ and $B$ are fixed in advance. The problem, where 
the bijection is not fixed in advance (see Figure \ref{ljoint-1}(i)), has been
studied by Aichholzer et al. \cite{aahk-tct-2003}.  
In this paper, we consider the other problem, where
the bijection is fixed in advance (see Figure \ref{ljoint-1}(ii)).

\medskip

Let $A=\{a_1, a_2, \ldots, a_n\}$ and $B=\{b_1, b_2, \ldots, b_n\}$ be two
disjoint sets of points in the plane, specified by their respective $x$ and $y$
coordinates.  A line segment $b_ib_j$ is called the {\it corresponding line segment} 
of the line segment $a_ia_j$ and vice versa. Similarly, 
a triangle $b_ib_jb_k$ is called 
the {\it corresponding triangle} of $a_ia_ja_k$ and vice versa. 
Let $\mathcal{T}(A)$ and $\mathcal{T}(B)$ denote the set of all triangulations 
of $A$ and $B$. The problem of joint triangulation of $A$ and $B$, as stated earlier,
is to find triangulations 
$T(A) \in \mathcal{T}(A)$ and $T(B) \in \mathcal{T}(B)$, if they exist, such that for each 
region bounded by a triangle $a_ia_ja_k$ in $T(A)$,  the corresponding triangle $b_ib_jb_k$ 
bounds a region in $T(B)$ 
(see Figure \ref{ljoint-1}(ii)).
The problem  was posed in 1987 by Saalfeld~\cite{saal-jttm-87}, and since then,
several researchers have worked on this problem  but the problem is still open.


\medskip

\begin{figure}[h]
\begin{center}
\centerline{\hbox{\psfig{figure=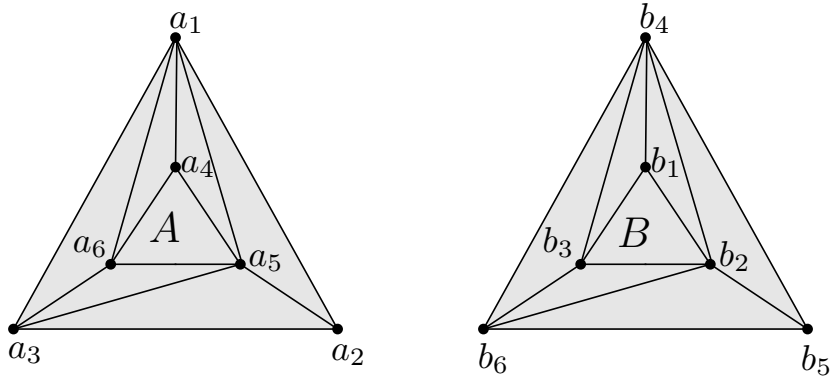,width=0.65\hsize}}}
\caption{The triangle $a_4a_5a_6$ does not contain any point of $A$,
whereas the corresponding triangle $b_4b_5b_6$ contains points of $B$.}
\label{ljoint2}
\end{center}
\end{figure}

The above definition of a joint triangulation of $A$ and $B$ needs some clarification. 
Consider triangulations $T(A)$ and $T(B)$ of point sets $A$ and $B$ respectively,
shown in Figure \ref{ljoint2}. It can be seen that for every line segment $a_ia_j$ in $T(A)$, 
the corresponding line segment $b_ib_j$ is in $T(B)$ and vice versa. However, the triangle 
$a_4a_5a_6$ does not contain any point of $A$, whereas the corresponding triangle $b_4b_5b_6$ 
contains points of $B$. Thus the triangles bounding the regions are different and we do not
consider this to be a joint triangulation.
This gives rise to the definition of a component triangle as defined by
Saalfeld  \cite{saal-jttm-87}. A triangle in $T(A)$ or $T(B)$ is said to be 
a {\it component triangle} of the triangulation if it does not contain any point in its interior.
Note that a triangle formed by three collinear points in $A$ or $B$ contains the middle point as its interior
and therefore, such a triangle is not considered as a component triangle.
Therefore, the problem of joint triangulation of $A$ and $B$ is to compute $T(A)$ and $T(B)$, if they exist,
such that a triangle $a_ia_ja_k$ is a component triangle in $T(A)$ 
if and only if the corresponding triangle 
$b_ib_jb_k$ is a component triangle in $T(B)$.

\medskip

In the next section, we propose two necessary conditions for this problem and
conjecture that they are sufficient. We also present an $O(n^3)$ time
algorithm for testing these necessary conditions. 
If the given set of points $A$ and $B$ satisfy the two necessary conditions,
we propose a greedy algorithm for constructing joint triangulations of
$A$ and $B$ in Section 3. The proposed algorithm has been implemented 
and experimental results suggest that the algorithms correctly construct joint 
triangulations of $A$ and $B$ whenever $A$ and $B$ satisfy the two necessary conditions.
Like two sets of points, a joint triangulation of two simple polygons
of same number of vertices can be defined analogously.
In Section 4, we present an
$O(n^3)$ time algorithm for computing a joint triangulation of two simple polygons
of $n$ vertices.
In Section 5, we conclude the paper with a few remarks.

\section{Necessary conditions}

Let $CH(A)$ and $CH(B)$ denote the boundary of convex hulls of $A$ and $B$ respectively.
We state the first necessary condition for the existence of a joint triangulation of $A$ and $B$,
which relates the edges of $CH(A)$ and $CH(B)$,

\medskip

\noindent {\bf Necessary condition 1:} 
If there exists a joint triangulation of $A$ and $B$, then $a_ia_j$ is an edge of $CH(A)$
if and only if the corresponding edge $b_ib_j$ is an edge of $CH(B)$.

\medskip

\noindent {\bf Proof:} Assume on the contrary that there is a joint triangulation of
$A$ and $B$ and an edge $a_ia_j$ is an edge in $CH(A)$ but the 
corresponding edge $b_ib_j$ is not an edge in $CH(B)$. Since $a_ia_j$ is an edge
of $CH(A)$, there exists only one component triangle (say, $a_ia_ja_k$) with $a_ia_j$ as an edge,
in any triangulation of $A$. On the other hand, we know that any joint triangulation must include 
$b_ib_j$ in the triangulation of $B$.
Since $b_ib_j$ is not an edge in $CH(B)$ by assumption, there are two component triangles 
(say, $b_ib_jb_k$ and $b_ib_jb_l$) with  $b_ib_j$ as an edge, in the triangulation of $B$. 
Since the component triangle $a_ia_ja_l$ is not present in the triangulation of $A$, 
this  contradicts the definition of a joint triangulation. 
\epf

\medskip

\begin{figure}
\begin{center}
\centerline{\hbox{\psfig{figure=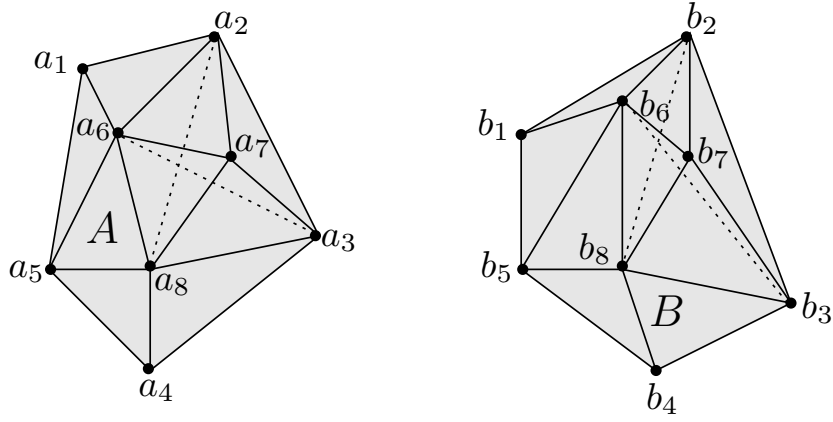,width=0.65\hsize}}}
\caption{On the edge $a_6a_7$, $a_6a_7a_8$ and $a_6a_7a_2$  are successor triangles.
 The corresponding triangles $b_6b_7b_8$ and $b_6b_7b_2$
 are also successor triangles on the edge $b_6b_7$}
\label{ljoint3}
\end{center}
\end{figure}

A triangle $a_ia_ja_k$ is said to be an {\it empty triangle} in $A$ if it does not contain
any point of $A$ in its interior. Let $S_A$ denote the set of all empty triangles in
$A$ whose corresponding triangles in $B$ are empty triangles in $B$. Let $S_B$ be the set
of triangles corresponding to the triangles in $S_A$.
It follows from the definition of a joint triangulation that only triangles 
from $S_A$ and $S_B$ can be component triangles in a joint triangulation of $A$ and $B$.
Let $a_ia_ja_k$ and $a_ia_ja_l$ be two triangles in $S_A$ such that they
lie on opposite sides of their common edge $a_ia_j$.  If $b_ib_jb_k$ and
$b_ib_jb_l$ also lie on opposite sides of their common edge $b_ib_j$,
then $a_ia_ja_l$ is a called a {\it successor triangle} of $a_ia_ja_k$ on the edge $a_ia_j$  
and vice versa. Analogously, $b_ib_jb_l$ is also called a {\it successor triangle} of $b_ib_jb_k$
on the edge $b_ib_j$ and vice versa.
In Figure \ref{ljoint3}, $a_6a_7a_8$ and $a_6a_7a_2$  are successor triangles on the edge
$a_6a_7$ and their corresponding triangles $b_6b_7b_8$ and $b_6b_7b_2$ are also successor triangles 
on the edge $b_6b_7$. On the other hand, $a_6a_7a_8$ and $a_2a_7a_8$ are not successor 
triangles on the edge $a_7a_8$ as $a_2$ and $a_6$ lie on the same side of $a_7a_8$.
Since successors of $a_ia_ja_k$ and $b_ib_jb_k$ are defined jointly,
in what follows, we say that $ijl$ is a successor triangle  of $ijk$ on edge
$ij$ and vice versa. Observe that $ijk$ can have more than one successor triangle on
an edge $ij$.  In Figure \ref{ljoint3}, $(2,6,8)$, $(7,6,8)$ and $(3,6,8)$ 
are successor triangles of $(5,6,8)$ on the edge $(6,8)$. It is obvious that
there is no successor triangle on any edge of the convex hull. 

\medskip

Intuitively, if a triangle $ijk$ is a component triangle in a joint triangulation, one of the
successors on each edge of $ijk$ that is not a convex hull edge is also a component triangle
in the joint triangulation. 
Let $S$ denote the maximal subset of triangles in $S_A$ and $S_B$ such
that each triangle $ijk$ in $S$ has at least one successor triangle in $S$, on the edges
$ij$, $jk$ and $ki$ that are not convex hull edges.  Note that if a triangle $ijk$ does not 
have a successor triangle on a non convex hull edge, then $ijk$ can not belong to $S$. 
We call triangles in $S$ as {\it legal triangles} and $S$ is called the
set of legal triangles. Now, we state the second necessary condition.

\medskip

\noindent {\bf Necessary condition 2:}
If there exits a joint triangulation of $A$ and $B$, then the set of legal
triangles $S$ is not empty.

\medskip

\noindent {\bf Proof:} If there is a joint triangulation of $A$ and $B$, then every 
component triangle in the joint triangulation has a successor triangle on each its non 
convex hull edges. So, every component triangle in a joint triangulation is a legal
triangle and hence, the set of legal triangles $S$ is not empty. \epf

\medskip
\noindent {\bf Conjecture:} There exists a joint triangulation of $A$ and $B$ if and only
if $A$ and $B$ satisfy the two necessary conditions.

\medskip

Let us present an algorithm for testing the necessary conditions. 
The first necessary condition can be tested by traversing the boundary of
the convex hulls of $A$ and $B$. Since the convex hulls  can be computed in 
$O(n \log n)$ time \cite{bkos-cgaa-97, ps-cg-85}, the first necessary condition can be 
tested in $O(n \log n)$ time.

\medskip

For testing the second necessary condition,  the algorithm starts by computing
all empty triangles in $A$ and $B$.
It has been shown by Dobkin et. al \cite{deo-secp-90} that all empty triangles in
a set of $n$ points in a plane can be computed in time proportional to the number of 
empty triangles which can be at most $O(n^3)$. So, $S_A$ and $S_B$ can be computed 
in $O(n^3)$ time. 
For every non-convex hull edge $ij$ of all triangles in $S_A$ and $S_B$, 
the algorithm checks whether there exists two triangles $ijk$ and $ijl$ on the edge $ij$
in $S_A$ as well as in $S_B$
such that $k$ and $l$ lie on opposite sides of $ij$ in both $A$ and $B$.
If $ij$ satisfies this condition, then there are successor triangles on the edge $ij$.
Otherwise, all triangles in $S_A$ and $S_B$ with $ij$ as an edge are removed from
$S_A$ and $S_B$, and the remaining two edges of every  deleted triangle are pushed into
a queue $Q$. For each edge $ef$ in $Q$, check whether there are successor triangles
on $ef$. If the condition is satisfied, then $ef$ is removed from the queue. Otherwise,
all triangles in $S_A$ and $S_B$ with $ef$ as an edge are removed from
$S_A$ and $S_B$, and the remaining edges of every deleted triangles are pushed into
the queue $Q$. This process is repeated till either $S_A$ and $S_B$ become empty or
the queue becomes empty. In the latter case, all remaining triangles in $S_A$ and $S_B$ have
successors on all non-convex hull edges, in which case they form the set of legal 
triangles $S$. Note that that the cost of processing edges in $Q$ can be assigned
to deleted triangles which can be at most $O(n^3)$. We state the result in the following
theorem.

\medskip


\noindent {\bf Theorem 1:}
Given two sets $A$ and $B$ of $n$ points in the plane, the two necessary conditions for a joint 
triangulation of $A$ and $B$  can be tested in 
$O(n^3)$ time.
 

\section{An algorithm for constructing joint triangulations}

In this section, we present two algorithms for finding a joint triangulation of 
$A$ and $B$ which run in $O(n^3)$ time. 
We assume that the set of legal triangles $S$ has been computed by 
the algorithm as mentioned in the previous section.
If the set $S$ is empty, clearly no joint triangulation exists.
So, we consider the other case when  $S$ is not empty.



\medskip

Constructing a joint triangulation of $A$ and $B$ involves finding a subset $T$ of legal triangles 
in $S$ 
forming a  triangulation in $A$ and the corresponding triangulation in $B$.
The algorithm uses a greedy method to obtain $T$. Initialize $S' = S$ and $T = \emptyset$. Take
any triangle $ijk$ from $S'$, add it to $T$ and delete all triangles in $S'$ that intersect 
the interior of the triangle $ijk$ in either $A$ or $B$.  
Repeat this process until $S'$ becomes empty. 
Our claim is that the triangles in $T$ form
a joint triangulation of $A$ and $B$. We have been unable to prove this
claim, which would also prove the sufficiency of the two necessary conditions. 
On the other hand, we have observed experimentally that whenever $S$ is not empty, 
the algorithm always 
finds a joint triangulation of $A$ and $B$. Readers may use our software for experimentation,
which is available at (http://www.tcs.tifr.res.in/$\sim$ghosh/Joint-triangulation/joint-triangulation.html).






 
$~~$
\section{Computing a joint triangulation of two simple polygons}
$~~$

\begin{figure}
\begin{center}
\centerline{\hbox{\psfig{figure=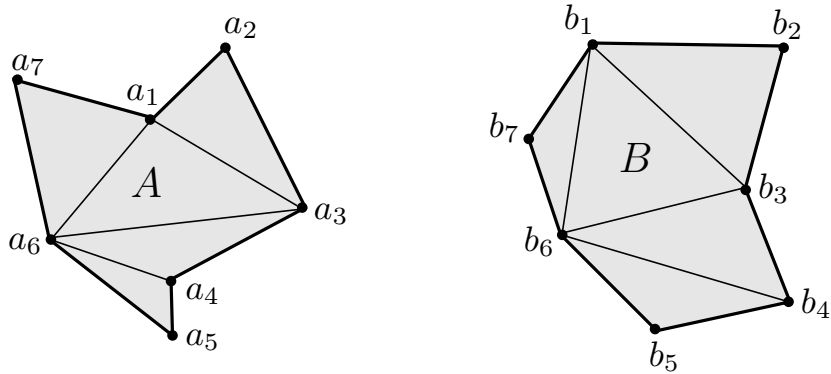,width=0.65\hsize}}}
\caption{A joint triangulation of two simple polygons $A$ and $B$.}
\label{ljoint4}
\end{center}
\end{figure}

In this section,  we present an
$O(n^3)$ time algorithm for computing a joint triangulation of two simple polygons
$A=(a_1, a_2, \ldots, a_n)$ and $B=(b_1, b_2, \ldots, b_n)$ using dynamic programming.
Two points $u$ and $v$ in a simple polygon are said to be {\it visible} if the line segment
$uv$ lies totally inside the polygon.  Let $VG(A)$ denote the visibility graph of $A$,
where vertices of $A$ are vertices of  $VG(A)$ and two vertices in $VG(A)$ are connected
by an edge if and only if the corresponding vertices in $A$ are visible in
$A$ \cite{s-vaip-2007}. The visibility graph $VG(B)$ of $B$ is defined analogously. 
We have the following observation (see Figure \ref{ljoint4}). 

\bigskip


\noindent {\bf Lemma 1:}
All edges of the triangles in a joint triangulation of  $A$ and $B$ must belong to
$VG(A)$ and $VG(B)$ respectively.

\bigskip

Let $IVG(A)$ denote the sub-graph of $VG(A)$ such that an edge $a_ia_j$ of $VG(A)$
belongs to $IVG(A)$ if and only if $b_ib_j$ is an edge of $VG(B)$.
Analogously, we define $IVG(B)$. It follows from Lemma 1 that we have to consider
only the edges of $IVG(A)$ and $IVG(B)$ in a joint triangulation of $A$ and $B$.
Since the visibility graph of a simple polygon can be 
computed in time proportional to the number of edges in the visibility graph,
which can be at most $O(n^2)$ \cite{Hershberger89}, $IVG(A)$ and $IVG(B)$ can be computed in
$O(n^2)$ time.

\medskip

\begin{figure}
\begin{center}
\centerline{\hbox{\psfig{figure=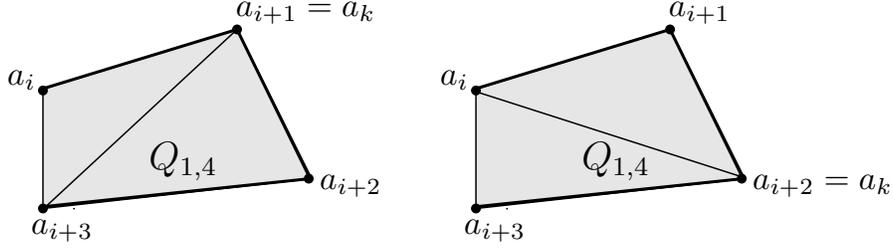,width=0.7\hsize}}}
\caption{Testing the sub-polygon $Q_{1,4}$ for a joint triangulation.}
\label{ljoint5}
\end{center}
\end{figure}

Let $SUB(A)$ denote the set of all sub-polygons of $A$ (including $A$ itself) that can
be formed 
by cutting $A$ using  only one diagonal  of $IVG(A)$. So, the size of sub-polygons 
in $SUB(A)$ varies from $3$ to $n$. We use a boolean function $M(Q)$ to indicate
whether a sub-polygon $Q$ admits joint triangulation. Since all sub-polygons 
of three vertices in $SUB(A)$ (say, $Q_{1,3}, Q_{2,3}, \ldots$) admit joint triangulations 
as they are triangles, 
$M(Q_{1,3}), M(Q_{2,3}), \ldots$ are set to be true.
Then the procedure considers sub-polygons $Q_{1,4}, Q_{2,4}, \ldots$ of $SUB(A)$ 
having four vertices. 
Let $Q_{1,4}=(a_i, a_{i+1}, a_{i+2}, a_{i+3})$ (see Figure \ref{ljoint5}).  
So, $a_ia_{i+3}$ 
is  the diagonal of $IVG(A)$ used to cut $A$ to form $Q_{1,4}$. Let $a_k$ be a vertex
of $Q_{1,4}$ such that edges $a_ia_k$ and $a_ka_{i+3}$ belong to $IVG(A)$. If no such
$v_k$ exists, then set  $M(Q_{1,4})$ to false. If  $a_{i+1}=a_k$ and
the triangle $(a_{i+1}, a_{i+2}, a_{i+3})$ admits triangulation found in the 
previous step, then set  $M(Q_{1,4})$ to true. If  $a_{i+2}=a_k$ and
the triangle $(a_i, a_{i+1}, a_{i+2})$ admits triangulation found in the 
previous step, then set  $M(Q_{1,4})$ to true. Otherwise, set  $M(Q_{1,4})$ to false.

\medskip

Similarly, the procedure considers sub-polygons $Q_{1,5}, Q_{2,5}, \ldots$ of $SUB(A)$ 
having five vertices by locating all possible such vertices $a_k$. This process is repeated
till the sub-polygon of size $n$ (i.e., $A$) is considered. In the following,
we state the major steps of the procedure.

\bigskip

\noindent {\bf Step 1:} Divide $A$ into sub-polygons using diagonals of $IVG(A)$ to form $SUB(A)$;

\medskip

\noindent {\bf Step 2:} Consider each edge of $A$ as a degenerated triangle; {\it For} each edge $a_ia_{i+1}$ {\it do}\\  $~~~~~~~~~~~~$$M(a_ia_{i+1}):=true$;

\medskip

\noindent {\bf Step 3:} {\it For} each sub-polygon $Q_{j,3}$ of size three in $SUB(A)$ 
{\it do} $M(Q_{j,3}):=true$; $size:=4$;

\medskip

\noindent {\bf Step 4:} {\it For} each sub-polygon $Q_{j,size}$ in $SUB(A)$ {\it do}

\medskip

{\bf Step 4.1:} {\it If} $Q_{j,size}=A$ {\it then} $i:=1$, $q:=n$, $k:=2$ and 
{\it goto} Step 4.3;
 
\medskip  

{\bf Step 4.2:} Let $a_ia_q$ be the diagonal used to cut $A$ to form
$Q_{j,size}=(a_i, a_{i+1}, \ldots, a_q)$;\\ $~~~~~~~~$$~~~~~~~~~~~~$$k:=i+1$; 

\medskip

{\bf Step 4.3:} {\it If} $a_ia_k$ and $a_qa_k$ are edges in  $IVG(A)$ and 
two sub-polygons formed by removing $~~~~~~~~$$~~~~~~~~~~~~$the triangle $(a_i,a_k,a_q)$ from
 $Q_{j,size}$ admit joint triangulations {\it then}\\ $~~~~~~~~$$~~~~~~~~~~~~$$M(Q_{j,size}):=true$;

{\bf Step 4.4:} {\it If} $k \neq q-1$ {\it then} $k:=k+1$ and {\it goto} Step 4.3;

\medskip

\noindent {\bf Step 5:} {\it If} $size \neq n$ then $size:=size+1$ and {\it goto} Step 4;

\medskip

\noindent {\bf Step 6:} {\it If} $M(A)$ is true {\it then} by backtracking identify diagonals of $IVG(A)$ giving  a joint\\ $~~~~~~~~~~~~~$triangulation {\it else} report that there is no joint triangulation.

\medskip

\noindent {\bf Step 7:}  Stop.

\bigskip

Since the procedure uses triangles formed by edges of $IVG(A)$ and $IVG(B)$, and
these triangles are added one at a time (i.e., $a_ia_ka_q$) to verify whether 
a joint triangulation exists for the sub-polygons formed by the union of triangles 
verified so far, the procedure correctly computes a joint triangulation
of $A$ and $B$ if it exists. Since the number of sub-polygons in $SUB(A)$ 
can be at most $O(n^2)$ and
the procedure can take $O(n)$ time for testing each sub-polygon, the overall time required
by the algorithm is $O(n^3)$. We state the result in the following theorem.

\medskip

\noindent {\bf Theorem 2:}
Given two simple polygons $A$ and $B$ of $n$ points, a joint triangulation of $A$ and $B$  
can be constructed in $O(n^3)$ time.

\section{Concluding remarks}

Let us mention some extensions of the basic problem. 
An immediate extension is to find a joint triangulation of $k$ sets of labeled points. 
It is easy to verify that for such a joint triangulation
to exist, boundary of the convex hulls of all sets of points must contain the same edges. 
Further, the notion of a
successor triangle can be extended to any number of sets of points in a natural way. 
A triangle $ijl$ is a successor of a triangle $ijk$ on the edge $ij$ if and only if 
it is a successor in all point sets. Thus we may define the set of legal triangles in 
an analogous way. We believe that the same conjecture holds for any number of sets of points.

\medskip

Further generalizations are possible by considering triangulations of objects other than 
just point sets. In particular, we can consider triangulations of any connected polygonal 
region with points and polygonal holes inside. The only difference here is that a triangle 
containing an edge of a hole boundary may not have a successor on that 
edge. Thus one necessary condition is that the hole boundaries must contain the same set 
of edges in all point sets. The definition of a successor triangle and a legal triangle 
may be modified accordingly, and the same algorithms can also be used. 
Again, we have observed empirically that if the set of legal triangles is not empty, 
there exists a joint triangulation, 
and it may be constructed in the same greedy fashion as for two point sets.

\medskip

If there is no joint triangulation of $A$ and $B$, it may still be possible
to obtain a joint triangulation by adding some points (say, $m$ Steiner points)
in $A$ and $B$. Naturally, it is desireable to add the smallest $m$ so that $A$ and $B$ admit
joint triangulation. Aichholzer et al. \cite{aahk-tct-2003} showed that it is always possible
to obtain joint triangulation of $A$ and $B$ (without a bijection)
by adding a linear number of Steiner points. One would expect a better
bound  where bijection between $A$ and $B$ is given in advance.
In the case of simple polygons $A$ and $B$ (without a bijection), Aronov et al. \cite{ass-ctsp-93}
showed that an addition of  quadratic number of  Steiner points is  sufficient and sometime necessary 
for constructing a joint triangulation. 

\medskip
\noindent {\large {\bf Acknowledgments}}

\medskip

The authors thank David Mount, Alan Saalfeld and Sudebkumar Pal for stimulating discussions.

\bibliographystyle{plain}
\bibliography{triangle}

\end{document}